\documentclass[twocolumn,notitlepage,nofootinbib,floatfix,superscriptaddress]{revtex4-1}
\usepackage{amsmath, amsfonts, amsthm, amssymb, bbm}
\usepackage[margin=1in]{geometry}
\usepackage{algorithm,algpseudocode}
\usepackage{graphicx,xcolor,tikz}
\usepackage[colorlinks=true, hyperindex, breaklinks, linkcolor=blue, urlcolor=blue, citecolor=blue]{hyperref}
\usepackage{physics}
\usepackage{qcircuit}

\newcommand{\unit}[1]{\,\mathrm{#1}}

\algblockdefx[ForEach]{ForEach}{EndForEach}[1]{\textbf{for each} #1 \textbf{do}}{\textbf{end for each}}

\graphicspath{{fewer_detections/figs/}}

\begin{document}

\preprint{APS/123-QED}

\title{Fast measurement of neutral atoms with a multi-atom gate}

\author{Yotam Vaknin}
\email{yotam.vaknin.qc@gmail.com}
\affiliation{Racah Institute of Physics, The Hebrew University of Jerusalem, Jerusalem 91904, Givat Ram, Israel}
\author{Ran Finkelstein}
\affiliation{School of Physics and Astronomy, Tel Aviv University, 69978, Israel}
\author{Ofer Firstenberg}
\affiliation{Department of Physics of Complex Systems, Weizmann Institute of Science, Rehovot 7610001, Israel}
\affiliation{Q-Factor, Tel Aviv 52510, Israel}
\author{Alex Retzker} 
\affiliation{Racah Institute of Physics, The Hebrew University of Jerusalem, Jerusalem 91904, Givat Ram, Israel}
\affiliation{AWS Center for Quantum Computing, Pasadena, CA 91125, USA}

\date{\today}

\begin{abstract}

Measurement time represents a critical bottleneck limiting the operational speed of neutral atom quantum computers, as it cannot be accelerated through parallelization like other quantum operations. We present a protocol for fast measurement of neutral atoms based on a new, fast multi-atom Rydberg gate that significantly reduces the measurement integration time and improves the measurement fidelity. Our approach employs a multi-qubit register of $N$ ancilla atoms within a single Rydberg blockade region to measure a single data qubit. This enables an $N$-fold enhancement in photon emission collections, while reducing the measurement's sensitivity to loss.  The scheme requires spectral separation between the data qubit and the
ancillae, achievable through either a dual-species architecture or a
targeted light shift. Beyond this, the scheme is straightforward to
implement: it relies only on global pulses, global photon collection, and avoids both atom shuttling and numerically optimized pulses.
Simulations of a
Cs--Rb platform demonstrate that with only five ancillae ($N=5$), measurement infidelity below $10^{-3}$ within $6 \unit{\mu s}$  is achievable. 
\end{abstract}
\maketitle

\section{Introduction}
\begin{figure*}[t]
    \centering

\begin{minipage}[t]{0.32\textwidth}
{\flushleft{\footnotesize (A)}\par}
  \centering
  \includegraphics[clip,trim={0 0 0 0},width=\linewidth]{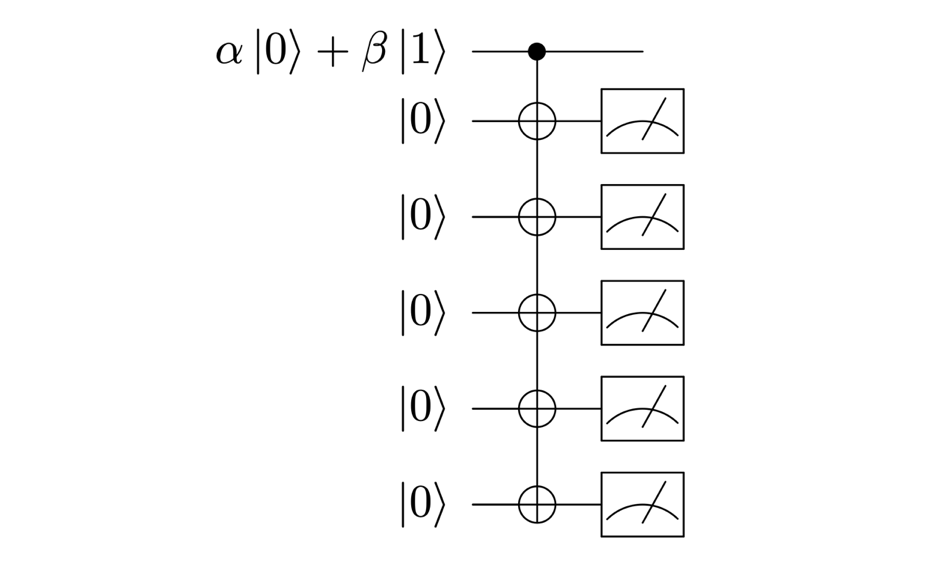}
\end{minipage}\hfill
\begin{minipage}[t]{0.32\textwidth}
{\flushleft{\footnotesize (B)}\par}
\centering
  \includegraphics[clip,trim={0 0 0 0},width=\linewidth]{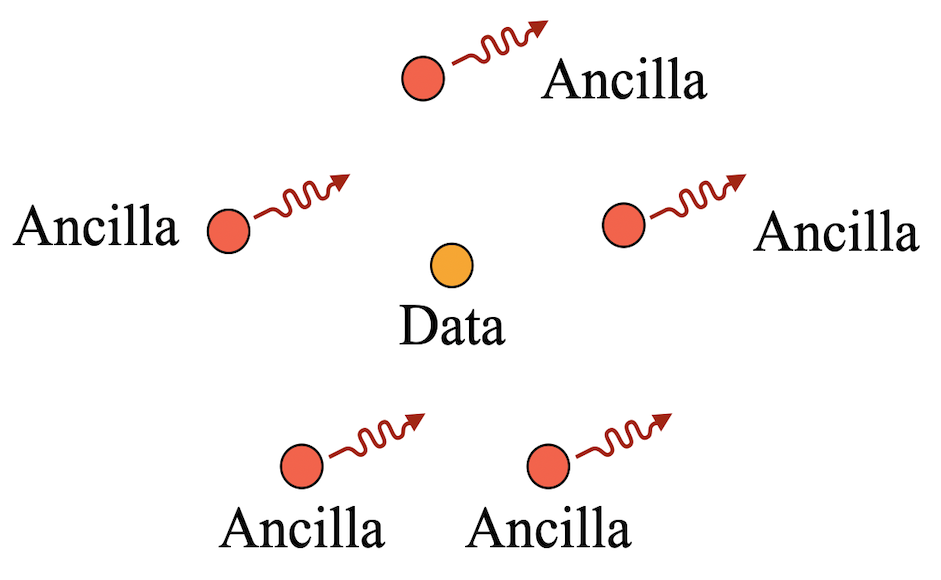}
\end{minipage}\hfill
\begin{minipage}[t]{0.32\textwidth}
{\flushleft{\footnotesize (C)}\par}
\centering
  \includegraphics[clip,trim={0 0 0 0},width=\linewidth]{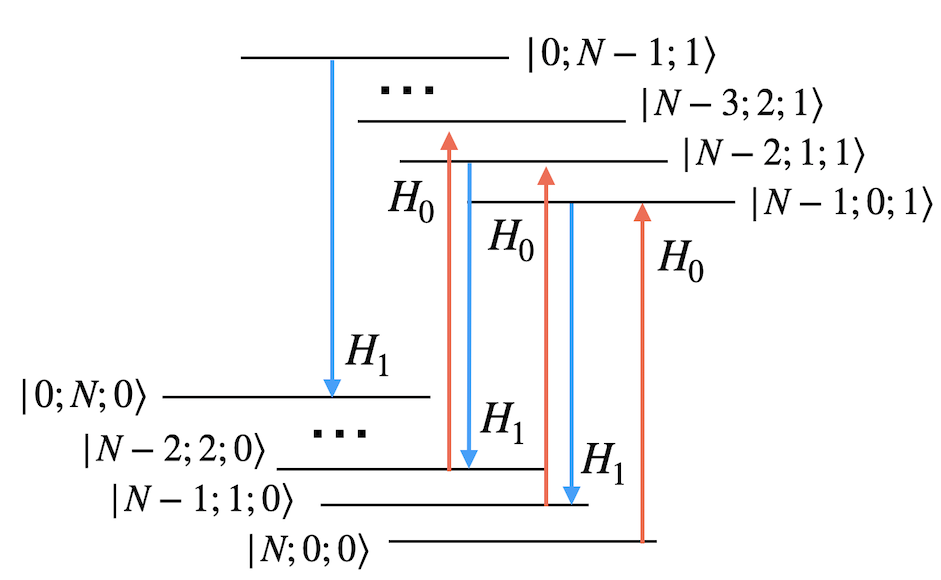}
\end{minipage}
  
    \caption{\textbf{Protocol for multi-atom gate and measurement}. (A) Gate level description of the protocol. A single atom state is copied to $N$  ancillae which are then measured at the same time. (B) A central data qubit is measured using $N=5$ photon emitting ancillae. The whole system is within a single Rydberg blockade region. (C) Level diagram and pulse sequence composing the protocol when the data qubit is excited. 
    Each horizontal line denotes a symmetric state (Eq.~\ref{eq:collective_ancilla_state}) labeled by the collective state of the ancillae, while the dynamics are invariant under permutations of the ancilla atoms. 
    Red (blue) arrows indicate global \(H_{0}\) (\(H_{1}\)) pulses. 
    Under Rydberg blockade (\(n_{R}\in\{0,1\}\)), \(H_{0}\) drives \(|n_{0};n_{1};0\rangle \leftrightarrow |n_{0}\!-\!1;n_{1};1\rangle\) while \(H_{1}\) drives \(|n_{0};n_{1};1\rangle \leftrightarrow |n_{0};n_{1}\!+\!1;0\rangle\). 
    Repeating the alternating sequence transfers one atom per pulse pair from \(|0\rangle\) to \(|1\rangle\), implementing the mapping \(|0\rangle_N \rightarrow |1\rangle_N\) in \(2N\) steps and realizing the state copy used for fast collective measurement.}
    \label{fig:levels}
\end{figure*}

Atom-based quantum computing has shown great promise in recent years, demonstrating the ability to control a large number of qubits, achieve high-fidelity gates, and maintain fidelities below the error-correcting threshold \cite{levine2019parallel,evered2023high,bluvstein2024logical,bluvstein2025architectural,muniz2025repeated,rines2025demonstration}. However, compared to other modalities, atom-based systems tend to have slower operation times, particularly in shuttling qubits and performing measurements. 

The large qubit count of atom-based computers \cite{manetsch2024tweezer,chiu2025continuous,holman2026trapping} could help mitigate some of this slowness through parallelization \cite{fowler2012time}, but it cannot make the \textit{clock speed} of the quantum computer faster than the measurement time \cite{fowler2012time}. This makes the measurement time a key metric limiting the speed of quantum computation, a limitation that is not imposed by the rate of initialization, shuttling, single- and two-qubit gates, or even the rate of magic state generation. Therefore,  reducing the measurement time in atom-based computers is uniquely important in making them a competitive modality for quantum computation.

The measurement time for neutral atoms is limited by their decay rate from the photon-emitting state. In the absence of an optical cavity \cite{goldwin2011fast} and without reducing the survival probability \cite{finkelstein2024universal, su2025fast} (which limits the measurement fidelity), acquiring enough fluorescence typically requires timescales on the order of milliseconds. This measurement time can be reduced by optically interacting with multiple atoms simultaneously \cite{PhysRevLett.99.120502,PhysRevLett.102.170502,PhysRevA.88.010303,zhang2025dualtypedualelementatomarrays,machu2025nondestructiveopticalreadoutmanipulation, PhysRevLett.119.180504,šumarac2026controllingrydbergatompolaritoninteractions}. In this work, we demonstrate how multiple atoms held in very close traps can be used to increase the photon flux using coherence excitation. This significantly reduces the integration time and improves the measurement fidelity by reducing its sensitivity to loss.

Our protocol copies the state of a data qubit into a multi-qubit atomic register \cite{xu2021fast}, which is then measured. This transfer is realized via a new gate protocol that simultaneously copies the data onto all ancillae, improving over previous sequential approaches \cite{finkelstein2024universal, tsai2026gatebasedreadoutcoolingneutral}. Our construction is fully analytical and does not require modulation of the drive phase or amplitude, an advantage over previous schemes, which relied on numerically optimized pulses and careful calibration to demonstrate multi-qubit gates \cite{cao2024multi, kazemi2025multi,mohan2025parametrized}.

The fidelity of our protocol is limited by the trade-off between the lifetime and dipolar interaction strength of high-lying Rydberg levels. We simulate a detailed scenario where $N$ Cesium (Cs) atoms are employed to measure a single Rubidium (Rb) atom, and find an optimal balance between these factors. Our results show that the proposed protocol improves over multi-atom gates in both speed and complexity \cite{cao2024multi}. In particular, it achieves more than an $N$-fold reduction in measurement time, scaling with the number of ancilla atoms, while improving the measurement fidelity.

Importantly, our protocol does not require intermediate shuttling, and only uses global addressing \cite{isenhower2011multibit}. The same protocol can be used as a fast way to apply multiple controlled-not operations, all controlled by the same qubit, at the same time. This kind of action is especially useful for table lookup \cite{babbush2018encoding, gidney2019windowed}, a standard optimization technique for many quantum algorithms including Shor's algorithm \cite{gidney2025factor, zhou2025resource}.

\section{Protocol}

Consider a single trap containing $N$ atoms, or $N$ traps with a single atom each within the same Rydberg blockade. We use these $N$ atoms as an ancilla register to quickly measure a data qubit. By preparing all the ancilla atoms in the same state, conditioned on the data qubit state, we can infer the qubit state by measuring the ancilla collectively. The key is that by having multiple atoms emit photons simultaneously, we can achieve an enhanced emission rate, significantly reducing the measurement integration time while reducing the sensitivity to atom loss.

We consider an implementation where the ancilla register is encoded in a different atomic species than the data qubit, so that global pulses act exclusively on the ancillae. As an alternative, one can apply a light shift to the data qubit to detune its Rydberg transition, which provides equivalent selectivity even with a single species. In the following, we focus on the two-species configuration, although the two approaches are fully interchangeable. 
  
 For a collective measurement, one first excites the data qubit to the Rydberg state conditional on its logical state, and then attempts to excite the ancilla atoms via the Rydberg level; this second step only succeeds if the data qubit was never excited. A simple protocol employing single-site addressing of the ancillae \cite{isenhower2011multibit}, in which each atom is excited separately, would require a total time scaling as $Nt$, where $t$ is the excitation time for a single atom. By instead using global pulses, we obtain the same effect in a time $T \approx t(2\sqrt{N} - \frac{1}{2})$, where $N$ denotes the number of ancilla atoms and $t$ is the duration of a $2\pi$ Rydberg pulse. In this scheme, all ancilla atoms are driven collectively through symmetric states, with atoms being successively excited until the entire ensemble is in the excited state (Fig.~\ref{fig:levels}.)

\subsection{Definitions }

We define two types of global pulses acting on the ancilla system:

\begin{align}
\label{eq:pulses}
H_0 &= \Omega\sum_i |0_i\rangle\langle R_i| + H_{\text{Blockade}} \\
H_1 &= \Omega\sum_i |1_i\rangle\langle R_i| + H_{\text{Blockade}}, \nonumber
\end{align}
where $\Omega$ is the drive Rabi frequency, $|0_i\rangle$ and $|1_i\rangle$ are the qubit states of the $i$-th ancilla atom, $|R_i\rangle$ is the Rydberg state, and $H_{\text{Blockade}}$ implements the Rydberg blockade interaction. Each Hamiltonian selectively drives its target transition, provided the qubit splitting satisfies $|\Omega| \ll E_{01}$, ensuring negligible off-resonant coupling to the other qubit state. Since the ancilla uses different atomic species from the data qubits, $H_0$ and $H_1$ affect only the ancilla atoms.

The system begins with the data qubit in an arbitrary superposition and all ancilla atoms initialized in the ground state:
$$|\psi\rangle = \alpha|0\rangle|0\rangle_N + \beta|1\rangle|0\rangle_N,$$
where $|0\rangle_N = \prod_{i=1}^N |0_i\rangle$ represents all N ancilla atoms in the ground state.

The ancilla state maintains complete symmetry under any permutation of the atoms throughout the protocol. We can therefore represent the state of just the ancilla $\ket{\psi_N}$ using three quantum numbers:
\begin{equation}
\label{eq:collective_ancilla_state}
    |\psi_N\rangle = |n_0; n_1; n_R\rangle,
\end{equation}
where $n_i$ is the number of atoms in state $i \in \{0, 1, R\}$. Due to the strong Rydberg blockade, $n_R \in \{0, 1\}$ always holds.

The pulse $H_0$ induces transitions between state pairs of the form:
$$|n_0; n_1; n_R\rangle \leftrightarrow |n_0 \pm 1; n_1; n_R \mp 1\rangle,$$
with $H_1$ acting similarly for $n_1$ and $n_R$. This creates a flip-flop mechanism between state pairs with bosonic amplification factors that we analyze below.

In this picture, we can rewrite the pulses $H_0$ and $H_1$ as interactions between three bosonic modes \cite{schwinger1952angular, arecchi1972atomic}:

\begin{align}
H_0 &= \Omega(t)(a_0^\dagger a_R + a_R^\dagger a_0) + \frac{\eta}{2}(a_R^\dagger)^2 a_R^2\\ 
H_1 &= \Omega(t)(a_1^\dagger a_R + a_R^\dagger a_1) + \frac{\eta}{2}(a_R^\dagger)^2 a_R^2, \nonumber
\end{align}
where $\Omega(t)$ is the drive energy, $\eta$ is the blockade energy and $a_i$ are the ladder operators for mode $i \in \{0, 1, R\}$. For completeness, we inlcuded a short derivation in Appendix~\ref{app:bosonic}. The symmetry between the ancilla qubits makes this description exact if the blockade energy is constant between pairs of atoms. Our analytical analysis will further assume the approximation of infinite blockade energy, although in practice, the atoms in the ancilla register would be distributed in some 2D array, and their blockade energy would not be uniform. Our simulations below analyze the effect of the geometry of the atoms on the fidelity of the measurement, by taking into account the different (finite) pair-wise blockade energies.

\subsection{ Pulse Sequence}

\begin{figure*}[t]
  \centering

  \begin{minipage}[t]{0.32\textwidth}
    
    \centering
    \flushleft{\footnotesize{(A)}}
    \includegraphics[clip,trim={0 0 0 0},width=\linewidth]{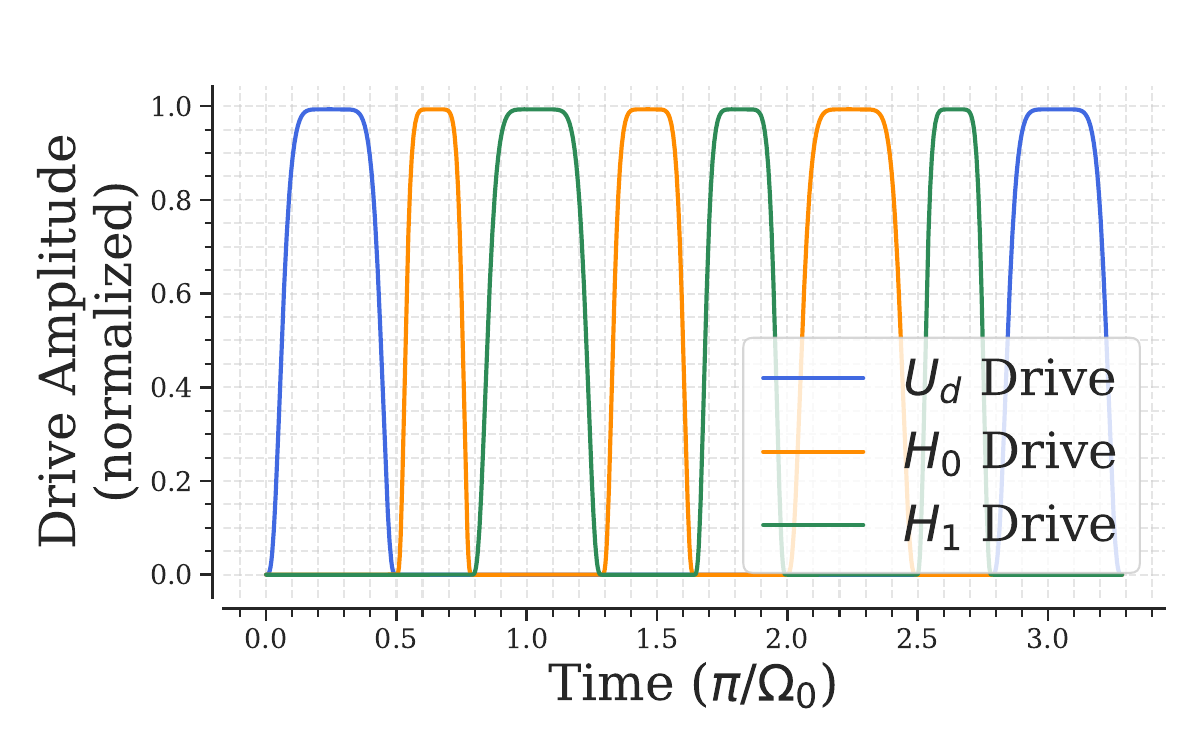}
    
  \end{minipage}\hfill
  \begin{minipage}[t]{0.64\textwidth}
  
    \centering
    \flushleft{\footnotesize{(B)}}
    \includegraphics[clip,trim={0 0 0 0},width=\linewidth]{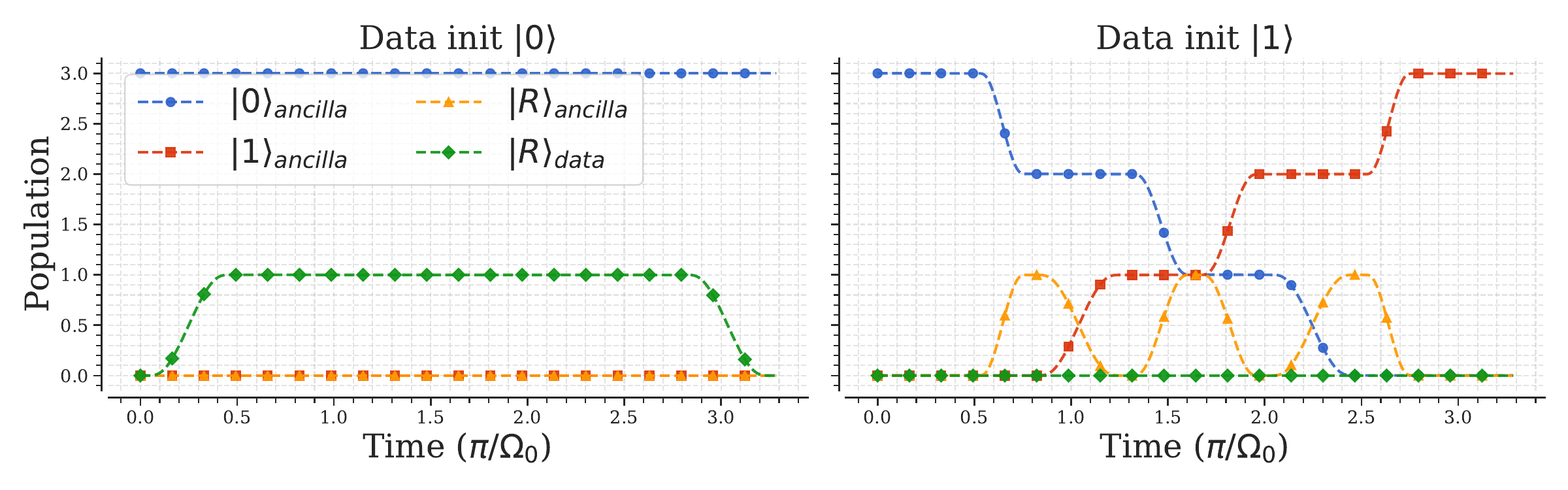}
  \end{minipage}\hfill

  \caption{\textbf{Drive sequence for $N=3$ ancilla atoms}.
  The multi-atom copying gate (multi-target $CX$) is composed of $2N+2$ pulses. The initial and final pulses $U_d$ excite and later return the single data qubit to and from the $\ket{R}$ state. The rest of the pulses are alternating $H_0, H_1$ drives that excite a single $\ket{0}$ state to the $\ket{1}$ state. (A) Drive amplitude of the 3 pulse types. (B) Population of the states $\ket{0},\ket{1},\ket{R}$  in the ancilla atoms and $\ket{R}$ in the data qubit during the gate, for both initial states of the data qubit $\ket{0}$ (left) and $\ket{1}$ (right). \label{fig:drive_sequence} }
\end{figure*}

We will now describe a pulse sequence that copies the state of the data qubit to the ancilla register. First, we apply a pulse $U_d$ on the data qubit that transforms the $|0\rangle$ state to the Rydberg state:
$$U_d|\psi\rangle = \alpha|R\rangle|0\rangle_N + \beta|1\rangle|0\rangle_N.$$
Due to the Rydberg blockade from the data qubit, the state $|R\rangle|0\rangle_N$ remains unaffected by subsequent $H_0$ and $H_1$ pulses. We therefore focus on finding a pulse sequence that implements the transformation $|1\rangle|0\rangle_N \to |1\rangle|1\rangle_N$, where $|1\rangle_N = \prod_{i=1}^N |1_i\rangle$.

We achieve this transformation by sequentially moving an atom from the $\ket{0}$ state to the $\ket{R}$ state using $H_0$, and then moving it from the $\ket{R}$ state to the $\ket{1}$ state using $H_1$. 

\begin{figure}[h]
  \centering
  \includegraphics[width=\linewidth]{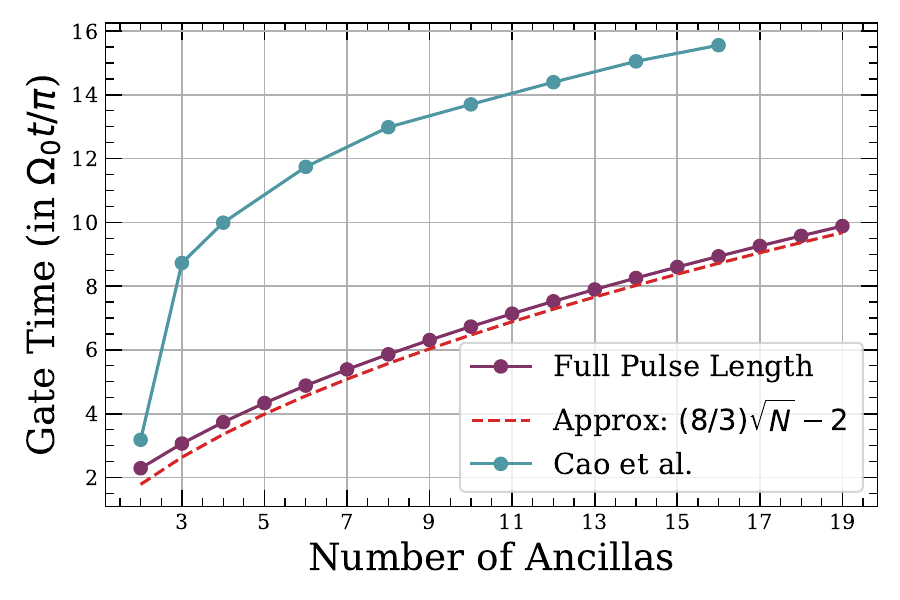}
  \caption{\textbf{Comparison of gate times as a function of the number of ancillae ($N$) --} Our sequence is compared with a similar construction optimized numerically~ \cite{cao2024multi}. To make the comparison fair, we increased the gate time by a factor of $4/3$ over Eq~\ref{eq:total_gate_time}, to account for the ramp-up and ramp-down of the drives. Gate times are measured in units of $\pi$ phases, i.e. $\Omega_0 t/\pi$. See Sec~\ref{sec:gate_sim} for details.}
\label{fig:gate_time_comparison}
\end{figure}

The pulse sequence induces the following state progression:
\begin{align*}
|N;0;0\rangle &\to |N-1;0;1\rangle \to |N-1;1;0\rangle \to  \\ 
&|N-2;1;1\rangle \cdots \to |0;N-1;1\rangle \to |0;N;0\rangle,
\end{align*}
which is achieved using the following set of transformations:
\begin{align}
U =& \exp(-iH_1 t_{2N})\exp(-iH_0 t_{2N-1})\cdots \\
  & \exp(-iH_1 t_2)\exp(-iH_0 t_1). \nonumber
\end{align}
The pulse durations are determined by the bosonic amplification factors. Before step $2k$, we have $n_0 = N-k$, $n_1 = k$, $n_R = 0$, and before step $2k+1$, we have $n_0 = N-k-1$, $n_1 = k$, $n_R = 1$. To achieve a single $\pi$ rotation each time, the pulse times would need to be:
$$t_i = \begin{cases}
\pi/\left(2\Omega\sqrt{N-k}\right) & \text{if } i = 2k \\
\pi/\left(2\Omega\sqrt{k+1}\right) & \text{if } i = 2k+1,
\end{cases}$$
The total time of these $2N$ pulses is therefore:

\begin{equation} 
\label{eq:gate_time approximation}
T_{2N} = \sum t_i = \frac{\pi}{\Omega}\sum_{n=1}^N \frac{1}{\sqrt{n}} \approx \frac{\pi}{2\Omega}(4\sqrt{N} - 3),
\end{equation}
after completing the pulse sequence and applying the inverse pulse $U_d^{-1}$ on the data qubit, the final state becomes:
\begin{equation}
\label{eq:final_state}
\alpha|0\rangle|0\rangle_N + \beta|1\rangle|1\rangle_N.
\end{equation}
Including the additional $\pi/\Omega$ time for the two $U_d$ pulses,  the total gate time is well approximated by 
\begin{equation}
\label{eq:total_gate_time}
T \approx \frac{\pi}{2\Omega}(4\sqrt{N} - 1).
\end{equation}
As shown in Fig~\ref{fig:gate_time_comparison}, the exact gate time matches this approximation, which is compared with an earlier multi-atom gate of Ref.~\cite{cao2024multi}. Despite being sequential, our protocol is substantially more efficient than the parallel implementation of Ref.~\cite{cao2024multi}.

The final state in Eq~\ref{eq:final_state}. demonstrates that all $N$ ancilla atoms are now prepared in a state that directly reflects the data qubit's logical state, enabling rapid collective measurement with enhanced photon emission. 


\section{Measurement fidelity}

The overall measurement fidelity is limited by two independent components: (i) the fidelity of the multi-qubit copying gate that maps the data qubit onto the ancilla registers, and (ii) the fidelity of the subsequent photon-scattering readout of the ancillae. In particular, the total measurement infidelity $IF$ cannot be lower than the gate infidelity $IF_{\text{gate}}$, which sets a fundamental floor on performance. We estimate both contributions using numerical simulations.

\subsection{Gate Simulation}
\label{sec:gate_sim}
The fidelity of our gate is limited by two competing effects: the short lifetime of the Rydberg state, and the finite interaction energy between the Rydberg atoms. Decay from the Rydberg state would scramble the protocol. Most significantly, if the data qubit decays from its Rydberg state during the protocol, our pulses would excite the ancilla for both logical states. This effect can be mitigated by increasing $\Omega$ such that the probability of decay from the Rydberg state becomes negligible, but $\Omega$ is limited by the blockade energy. If $\sqrt{N}\Omega$ is comparable to the Rydberg blockade energy, the Rydberg state would be doubly excited resulting in a different final state. In principle, the gate infidelity should include a component representing the dephasing between the two data-qubit states, but since our protocol concludes with a projective measurement, dephasing won't limit the final measurement fidelity (see Eq.~\ref{eq:total_variational_distance} below). 

The trade-off between the two sources of noise can be studied numerically, by modeling the atoms as a $4$ level system with the states $\ket{0},\ket{1},\ket{R}$ and a loss state. We introduce decay from the $\ket{R}$ state to the loss state and evaluated the blockade energy using standard techniques \cite{vsibalic2017arc}.

Similar to the experiment described in \cite{anand2024dual}, our simulation include $N$ Cs atoms that measure a single Rb atom. We evaluate the Rydberg blockade energy between the $\ket{77S_{1/2}}$ states of the Cs and the $\ket{78S_{1/2}}$ state of the Rb atom, but unlike \cite{anand2024dual} we use the van-der-Waal (vdW) interaction as the mechanism for the Rydberg blockade. We evalute the blockade energy and radiative lifetime of the Rydberg states using \cite{vsibalic2017arc}. The $C_6$ vdW term \cite{anand2024dual,vsibalic2017arc} between Cs atoms are $C_6^{Cs}=-2.9\unit{GHz\  \mu m^6}$ and between Cs and Rb, $C_6^{Cs-Rb}=-1.7\unit{GHz\  \mu m^6}$. We found room temperature lifetime of the Cs and Rb Rydberg states to be $T_1^{Cs}=176\unit{\mu s}$ and $T_1^{Rb} = 190 \unit{\mu s}$, respectively. The minimal distance between the traps in our simulation is $d=2 \unit{\mu m}$, above the LeRoy radius \cite{leroy1970dissociation}. The geometric distribution of the atoms is shown in Appendix~\ref{SI:gate_infidelity}. 

An important source of noise are non-adiabatic leakage to the detuned doubly-excited Rydberg states. Pulse shaping changes the scaling of leakage from $\left(\Omega/\Delta\right)^{2}$ to $\left(\Omega/\Delta\right)^{4}$ \cite{PhysRevA.88.013402}, where $\Delta$ is the detuning from the non-resonant transition. We avoid these non-adiabatic transitions by ramping the flip-flop term of $H_0$, $H_1$ and $U_d$ using the following function:
$$ 
\Omega\left(t\right)=\Omega_0\left(1-\left(1-\sin\left(\pi t/T\right)\right)^{4}\right)^{4},
$$
where $T$ is the total time of the pulse, and $\Omega$ is the drive frequency.  The total drive sequence for $N=3$ ancilla atoms is shown in Fig~\ref{fig:drive_sequence}.

After the interaction, a certain number of Cs atoms occupy the $\ket{1}$ state, with the corresponding probability distribution determined by the initial state of the Rb atom. We denote this distribution by $p_n^{\ket{0}}$ ($p_n^{\ket{1}}$), representing the probability of finding $n$ Cs atoms in $\ket{1}$ given that the Rb atom was initially prepared in $\ket{0}$ ($\ket{1}$). In the ideal case, these distributions satisfy $p_0^{\ket{0}} = 1$ and $p_N^{\ket{1}} = 1$, with all other probabilities vanishing. We use a statistical bound $IF_{\text{gate}}$ on the measurement infidelity ($IF$), since it is limited by our ability to differentiate between these two distributions $p_n^{\ket{1}}, p_n^{\ket{0}}$. We can bound it using their total variation distance ($\text{TVD}$):

\begin{equation}
\label{eq:total_variational_distance}
IF_\text{gate}=\frac{1}{2}-\frac{1}{2}\text{TVD}=\frac{1}{2}-\frac{1}{4}\sum_n\left|p_n^{\ket{0}}-p_n^{\ket{1}}\right|.
\end{equation} 

Note that this differs from the standard gate infidelity, which should account for dephasing of the final state. Because our protocol terminates in a projective measurement, the relevant figure of merit is the statistical distinguishability of the two distributions, which Eq.~\ref{eq:total_variational_distance} captures exactly.

\subsection{Measurement Simulation}

In our model, each ancilla is read out by driving a closed (cycling) optical transition for a duration $t_{\mathrm{meas}}$, producing a stream of detected photon counts. The camera registers only a small fraction of the fluorescence photons, and it also registers background counts originating from scattered light from the excitation laser. We model the resulting count statistics with a discrete-time Markov process with time step $dt$. In each step, an ancilla that remains trapped produces a detected fluorescence count with probability $dt/T_{\mathrm{photon}}$. Independently, a background count occurs with probability $dt/T_{\mathrm{noise}}$. Finally, each ancilla is lost from its trap with probability $dt/T_{\mathrm{loss}}$; once loss occurs, that ancilla contributes no further fluorescence counts for the remainder of the measurement window. The numerical values used in our simulation are given in the Appendix~\ref{SI:numerical_table}.

We label the number of observed photons $m$ for a Rb atom initilized in the state $\ket{0}$ ($\ket{1}$) as $q_m^{\ket{0}}$ ($q_m^{\ket{1}}$). Again, the measurement infidelity is limited by our ability to differentiate between the two distributions $q_m^{\ket{0}}$, $q_m^{\ket{1}}$:
\begin{equation}
\label{eq:total_variational_distance}
IF=\frac{1}{2}-\frac{1}{4}\sum_m\left|q_m^{\ket{0}}-q_m^{\ket{1}}\right|.
\end{equation} 

\begin{figure}[h]
  \centering
  \includegraphics[width=\linewidth]{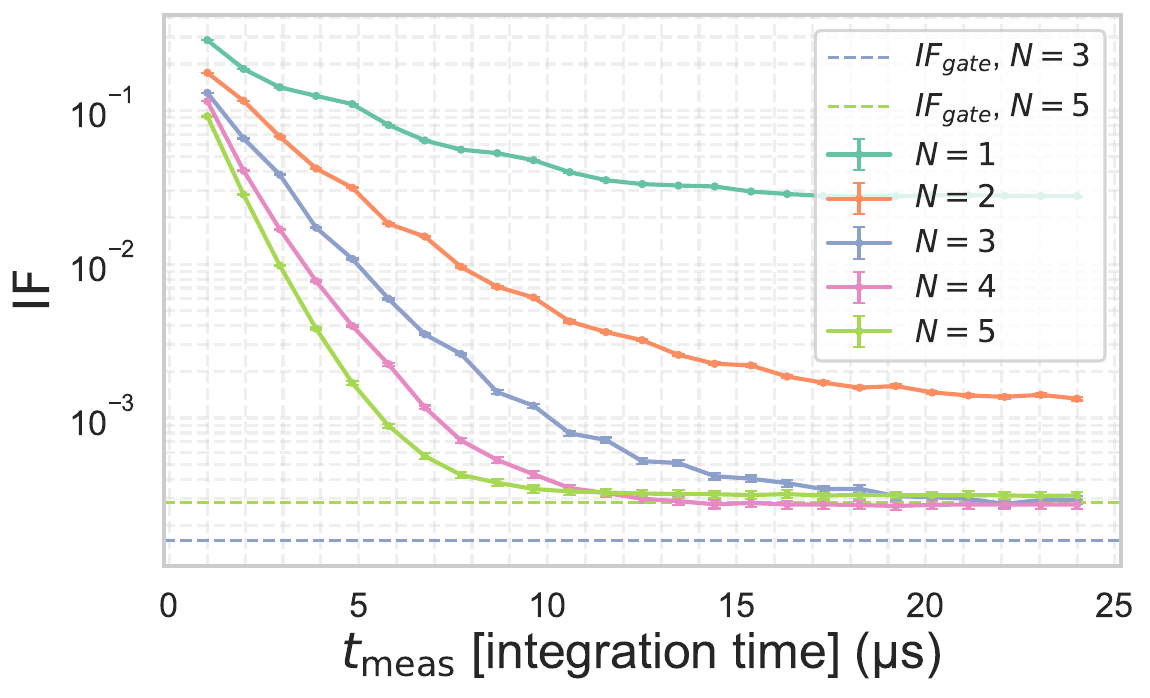}
\caption{\textbf{Measurement infidelity $\mathrm{IF}$ versus integration time $t_{\mathrm{meas}}$ for detecting the ancilla register} -- For each $N$, the data qubit is first mapped onto $N$ ancilla atoms via the copying gate, and the ancillae are then measured by collecting photons for an integration time $t_{\mathrm{meas}}$. The measurement infidelity is quantified via the total-variation distance between the relevant photon-count distributions, as defined in Eq.~\ref{eq:total_variational_distance}. The measurement model includes background photon counts and atom loss. Dashed horizontal lines indicate the corresponding gate-infidelity floor $\mathrm{IF}_{\text{gate}}$ (shown for $N=3$ and $N=5$), which sets the minimum achievable total readout error even in the long-integration-time limit. For comparison with standard readout, the $N=1$ case shows the result of a perfect gate followed by a noisy single ancilla detection.}
\label{fig:measurement_fidelity}
\end{figure}

For each number of ancilla atoms $N$, we chose the $\Omega$ that minimized $IF_{gate}$ in the range we simulated ($4-15 \text{MHz}$), see Appendix~\ref{SI:gate_infidelity}. We sampled the process of collecting photons for the given distribution of excited atoms $p_n^{\ket{0}},p_n^{\ket{1}}$. Because of background photons and high loss rate, the measurement fidelity is saturated around $25\mu s$. We show the measurement fidelity for each $N$ is figure~\ref{fig:measurement_fidelity}.

\begin{figure}
    \includegraphics[width=\linewidth]{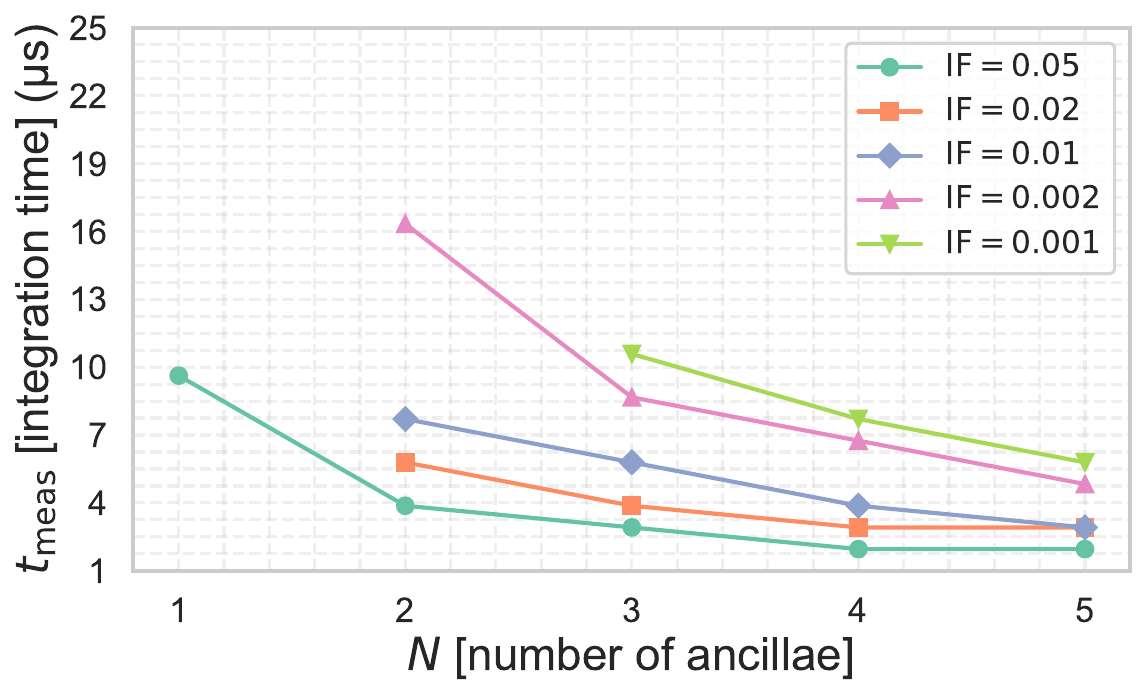}
   \caption{
       \textbf{Minimal measurement integration time required for target
       measurement infidelity $\mathrm{IF}$, as a function of  $N$} -- For each $N$ the optimal drive amplitude $\Omega$ is
       selected from Appendix~\ref{SI:gate_infidelity}, the data qubit is mapped
       onto the ancilla register via the copying gate, and photons are
       collected for a duration $t_\mathrm{meas}$ until the target infidelity
        is first reached. Missing points indicate values of $N$ for which the
        gate-infidelity floor $\mathrm{IF}$ (see
        Fig.~\ref{fig:measurement_fidelity}) already exceeds the target $\mathrm{IF}$,
        making it unreachable regardless of integration time.
        This floor is set by the finite trapping lifetime during readout.
    }
    \label{fig:min_meas_time}
\end{figure}

The results demonstrate a significant improvement in both measurement speed and fidelity. A single-atom measurement saturates at roughly $2\%$ infidelity, limited by atom loss during the detection window. With $N=5$ ancillae, the infidelity reaches $0.1\%$ in approximately $6\,\mu\mathrm{s}$, a faster measurement which simultaneously improving the infidelity by more than an order of magnitude. This improvement arises from two complementary effects: multiple emitting ancillae provide redundancy against atom loss together with a larger photon flux. These improvements are summarized more explicitly in Fig.~\ref{fig:min_meas_time}.

     In addition to the register-aggregated photon counting discussed above, we also considered an atom-resolved detection model in which the collection optics can distinguish which ancilla emitted each detected photon. In this case, the measurement record is a vector of per-atom photon counts rather than a single total count. Atom-resolved readout can, in principle, provide more information. The optimal decision rule for atom-resolved detection is no longer a one-dimensional threshold on the total count. Instead, use a maximum likelihood estimation on the likelihood of the observed measurement, conditioned on the initial state of the data qubit.  For the parameters relevant to our setup, we find that atom-resolved processing yields very similar fidelity compared with the aggregated-count model, with most of the variance coming from the finite sampling noise. Full details of the atom-resolved model are provided in Appendix~\ref{app:atom_resolved_readout}. We observed that atom-resolved detection can outperform aggregated counting in the regime of very low background noise and large separation between
$p_n^{\ket{0}}$ and $p_n^{\ket{1}}$ at $n \in \{1, N{-}1\}$, but
achieving this regime requires parameters outside the physically
realistic range considered here.

\section{Conclusion}
We presented a novel protocol for fast measurement of neutral atoms using multi-atom gates that achieves significant improvements in both measurement speed and fidelity. The enhanced photon emission rates both reduces the integration time, and makes the measurement more robust to atom loss during detection. We also showed that our protocol is significantly faster than previous demonstrations of multi-atoms gates.

The two leading quantum computing platforms -- superconducting qubits and neutral atom
arrays -- occupy distinct points in the architectural trade-off space. Superconducting qubits
offer fast gate and measurement operations but are largely restricted to local connectivity,
with long-range links requiring dedicated hardware. This constrains superconducting
architectures to codes with limited connectivity \cite{kitaev2003fault, gidney2023yoked, Bravyi_2024}, though their fast clock speed
enables efficient execution within those codes. Neutral atoms, by contrast, benefit from
reconfigurable, nearly arbitrary connectivity, enabling the use of high-rate codes such as
quantum LDPC codes \cite{xu2024constant, cain2026shor}. However, their clock speed is limited by slow shuttling and
measurement times. Compiling algorithms such as Shor's factoring algorithm for fast
execution requires operating in the reaction-time limit~\cite{fowler2012time, zhou2025resource}, where the clock speed is
set by the measurement and decoding time (i.e. the reaction time) rather than by gate operations or shuttling. This comes at a cost of increasing the number of qubits, a trade-off well suited to atom arrays. In this regime, our
protocol reduces the reaction time from ${\sim}\,1\,\unit{\mathrm{ms}}$ to
${\sim}\,10\,\unit{\mu\mathrm{s}}$ (including $4 \mu s$ decoding), yielding a ${\sim}100\times$ improvement in clock speed and,
correspondingly, in the total execution time of algorithms that are reaction-time limited. Applying our protocol to the reaction-limited compilation of \cite{zhou2025resource} would reduce the estimated execution time of 2048-bit RSA factoring from $\sim 4 $ days to $\sim 1$ hour.

Beyond measurement applications, this protocol serves as an efficient method for implementing multiple CX operations with a shared control qubit, a capability particularly valuable for windowed lookup \cite{gidney2019windowed}. The global addressing scheme and symmetric state progression make the protocol well-suited for implementation in current atom-based quantum computers. This avoids long sequential CX gates, or more complex schemes that generate GHZ ancillae \cite{zhou2025resource}.

The measurement time reduction achieved by this work addresses a fundamental limitation of atom-based quantum computers, potentially making them more competitive with other quantum computing modalities where fast measurement and feedback are essential for practical quantum computation.

\section{Acknowledgments}
We acknowledge the support of TWIN-Q Quantum computing center.
A. R. acknowledges the support of ISF, the Israel Innovation Authority  and the Schwartzmann university chair. This research was supported by Israel Science Foundation research grant (ISF’s No. 4098/25) and the Maimonides Fund’s Future Scientists Center.

\bibliography{bib}
\clearpage

\appendix

\appendix
\section{Bosonic mode representation}\label{app:bosonic}

In this appendix, we derive the bosonic form of the Hamiltonians $H_0$ and $H_1$ from their first-quantization description.

Since the Hamiltonians $H_0$ and $H_1$ are invariant under permutations of the ancilla atoms, the dynamics are confined to the symmetric subspace. A basis for this subspace is given by the occupation-number states:
\begin{equation}
    |n_0; n_1; n_R\rangle = \mathcal{N} \sum_{\pi \in S_N} \pi\!\left( |0\rangle^{\otimes n_0} \otimes |1\rangle^{\otimes n_1} \otimes |R\rangle^{\otimes n_R} \right),
\end{equation}
where $n_0 + n_1 + n_R = N$, $S_N$ is the symmetric group, and $\mathcal{N}=\sqrt{n_0! n_1!n_R!/N!}$ is a normalization factor.

We introduce bosonic ladder operators $a_i^\dagger, a_i$ for each internal state $i \in \{0, 1, R\}$, defined by their action on the occupation-number states:
\begin{equation}
    a_i^\dagger |n_0; n_1; n_R\rangle = \sqrt{n_i + 1}\, |n_0; n_i + 1; n_R\rangle,
\end{equation}
with $[a_i, a_j^\dagger] = \delta_{ij}$ and $a_i^\dagger a_i |n_0; n_1; n_R\rangle = n_i\, |n_0; n_1; n_R\rangle$. These satisfy the fixed particle-number constraint $a_0^\dagger a_0 + a_1^\dagger a_1 + a_R^\dagger a_R = N$.

Consider the flip-flop term of $H_0$, which acts on the $k$-th atom as $|0_k\rangle\langle R_k|$, summed over all atoms:
\begin{equation}
    V_0 = \Omega \sum_{k=1}^{N} |0_k\rangle\langle R_k| + \mathrm{H.c.}
\end{equation}
To evaluate the matrix element $\langle n_0+1; n_1; n_R-1| \, V_0 \, |n_0; n_1; n_R\rangle$, we note that each of the $n_R$ atoms in state $|R\rangle$ can be flipped to $|0\rangle$, contributing one term to the sum. The resulting state has an overlap with $|n_0+1; n_1; n_R-1\rangle$ that includes a ratio of normalization factors, yielding:
\begin{equation}
    \langle n_0+1; n_1; n_R-1| \, V_0 \, |n_0; n_1; n_R\rangle = \Omega\sqrt{n_R}\sqrt{n_0+1}.
\end{equation}
This is precisely the matrix element of $\Omega\, a_0^\dagger a_R + \mathrm{H.c.}$\, in the occupation-number basis. An identical calculation applies to the flip-flop term of $H_1$, giving $\Omega\, a_1^\dagger a_R + \mathrm{H.c.}$

The Rydberg blockade restricts $n_R \in \{0,1\}$. In the bosonic language, this constraint is enforced by the on-site interaction $(\eta/2)(a_R^\dagger)^2 a_R^2$, which assigns an energy cost $\eta$ to the doubly-excited state $n_R = 2$. In the limit $\eta \to \infty$, the doubly-excited Rydberg states are projected out, recovering the blockade condition.

\section{Gate Infidelity}
\label{SI:gate_infidelity}

\begin{figure}[h]
  \centering
  \includegraphics[width=\linewidth]{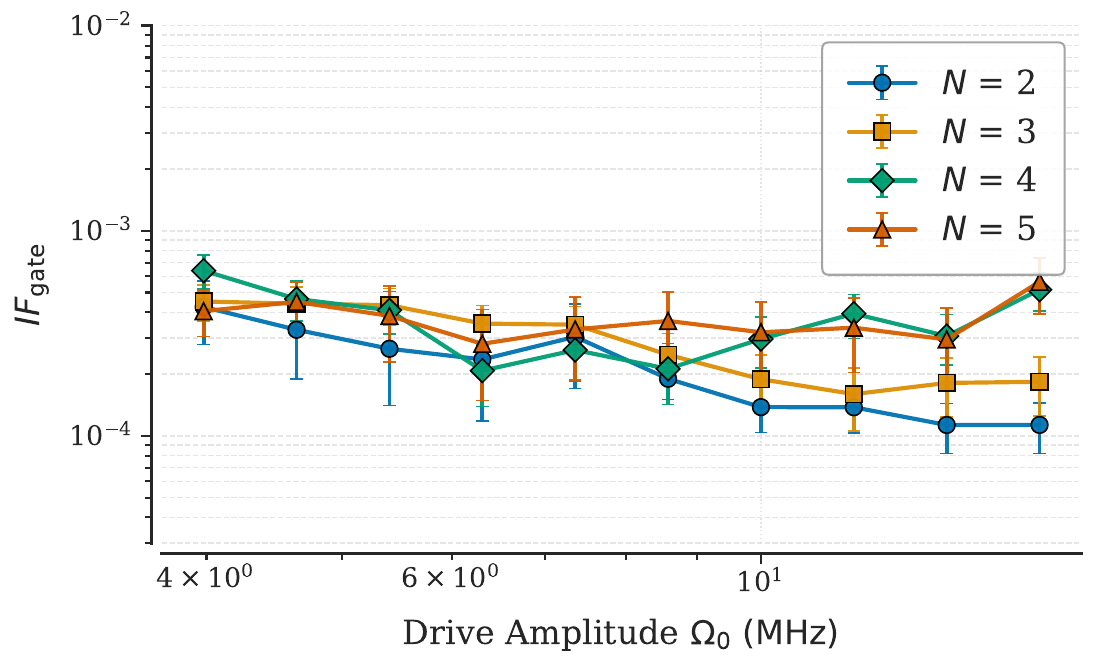}
  \caption{Monte-Carlo trajectory simulations of gate infidelity as a function of the Rabi drive amplitude $\Omega$ for the $N$-ancilla implementation. 
  The data qubit is a single Rb atom and the $N$ ancillae are Cs atoms. For each $N$, atoms are placed with a minimum separation of $2.2\,\mu\mathrm{m}$. 
  Pulse durations are scaled with $\Omega_0$ to preserve fixed pulse areas across the sweep. 
  The simulation includes finite Rydberg blockade and decay from the Rydberg manifold to a loss state. 
  Error bars show one standard deviation from finite trajectory sampling. 
  The infidelity $IF_{\mathrm{gate}}$ is computed from the total-variation-distance metric defined in Eq.~\ref{eq:total_variational_distance}.}
  \label{fig:gate_infidelity}
\end{figure}

\begin{figure*}
\centering
\includegraphics[width=\textwidth]{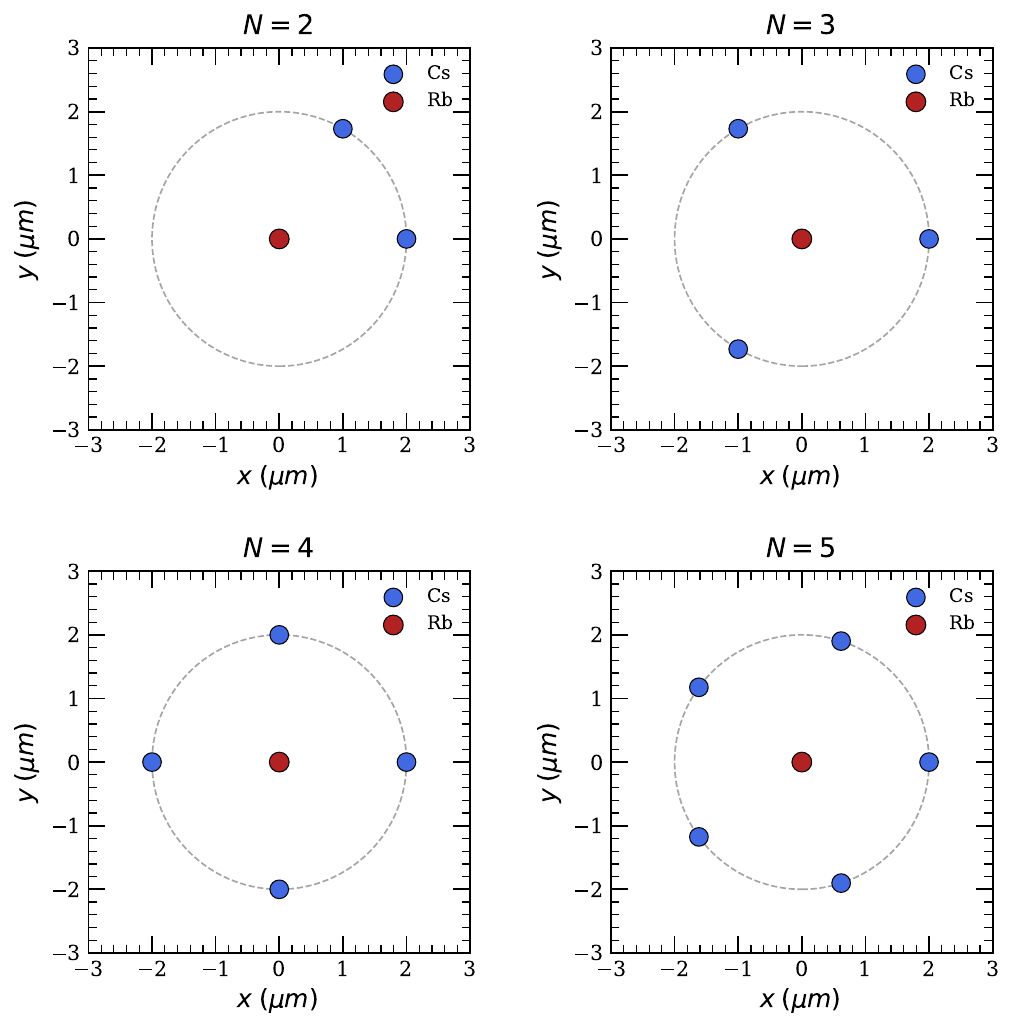}
    \caption{Spatial arrangement of the Cs ancilla atoms (blue) and the Rb data qubit (red) for $N = 2, 3, 4, 5$ ancillae. All atoms lie within a single Rydberg blockade region. The minimum inter-atom separation is $d = 2\,\mu$m. These geometries are used in the gate simulations, and define the non-uniform pairwise blockade energies.}
\label{fig:atomic_grid}
\end{figure*}

The measurement infidelity is limited by the gate infidelity $FI_{\text{gate}}$. The drive amplitude trades-off the two major sources of noise: decay from the Rydberg state, and leakage to doubly-excited rydberg states. The finite blockade energy is a function of the geometric distribution of the atoms, shown in Fig~\ref{fig:atomic_grid}. For lower $n$ Rydberg state, the trade off is more pronounce because of shorter lifetime and weaker blockade. Fig~\ref{fig:gate_infidelity} shows the average gate infidelity $IF_{gate}$ for different values of the the drive amplitude $\Omega_0$, with the error bars coming from finite trajectory sampling.

\section{Atom-resolved readout model and MLE classification}
\label{app:atom_resolved_readout}

In the atom-resolved detection scheme, we assume that the imaging system can
spatially resolve individual ancilla sites, so that detected photons are
attributed to specific ancilla locations rather than pooled into a single
register-wide count. The measurement record is then an $N$-dimensional vector
\begin{equation}
\mathbf{m} = (m_1,m_2,\dots,m_N),
\end{equation}
where $m_i$ is the number of photons detected at ancilla site $i$ during a
measurement window of duration $T$. In the following, we derive the optimal maximum-likelihood estimator for classifying
the data-qubit state from $\mathbf{m}$.

\subsection{Single-ancilla photon statistics}

We derive the photon-count distribution for a single ancilla location by considering two independent processes: fluorescence from the atom (if excited) and background photons.

An excited ancilla that remains trapped emits photons as a Poisson process with time constant $T_{\mathrm{photon}}$. The ancilla is lost at a random time governed by an exponential distribution with time constant $T_{\mathrm{loss}}$, so the probability density of loss at time $t$ is $p_{\mathrm{loss}}(t) = e^{-t/T_{\mathrm{loss}}}/T_{\mathrm{loss}}$, while the probability of surviving the full measurement window is $e^{-T/T_{\mathrm{loss}}}$. Background photons arrive as an independent Poisson process with time constant $T_{\mathrm{bg}}$ over the entire window $[0,T]$, regardless of the state of the atom.

The probability of detecting $n$ atom-fluorescence photons, marginalizing over the unknown loss time, is therefore
\begin{align}
P_{\mathrm{atom}}(n) &= \underbrace{\int_0^T \frac{e^{-t/T_{\mathrm{loss}}}}{T_{\mathrm{loss}}} \, e^{-t/T_{\mathrm{photon}}} \frac{(t/T_{\mathrm{photon}})^n}{n!}\, dt}_{\text{atom lost at }t\in[0,T]}\nonumber \\
&+ \underbrace{e^{-T/T_{\mathrm{loss}}} \, e^{-T/T_{\mathrm{photon}}} \frac{(T/T_{\mathrm{photon}})^n}{n!}}_{\text{atom survives to }T}.
\label{eq:P_atom}
\end{align}
Defining $\lambda = 1/T_{\mathrm{photon}} + 1/T_{\mathrm{loss}}$, the integral evaluates to
\begin{equation}
\frac{T_{\mathrm{photon}}}{T_{\mathrm{loss}}+T_{\mathrm{photon}}} \left(\frac{T_{\mathrm{loss}}}{T_{\mathrm{loss}}+T_{\mathrm{photon}}}\right)^{\!n} \frac{\gamma(n+1,\,\lambda T)}{n!},
\end{equation}
where $\gamma(a,x)$ is the lower incomplete gamma function. The background-photon distribution is simply Poisson:
\begin{equation}
P_{\mathrm{bg}}(m) = e^{-T/T_{\mathrm{bg}}} \frac{(T/T_{\mathrm{bg}})^m}{m!}.
\end{equation}

Since the atom-fluorescence and background processes are independent, the total detected-photon distribution at a location containing one atom is the convolution
\begin{equation}
P(n) = \sum_{k=0}^{n} P_{\mathrm{atom}}(k)\, P_{\mathrm{bg}}(n-k),
\label{eq:P_with_atom}
\end{equation}
while at a location with no excited atom, the distribution is purely background:
\begin{equation}
Q(n) = P_{\mathrm{bg}}(n).
\label{eq:Q_no_atom}
\end{equation}

\subsection{Multi-ancilla likelihood}

Conditional on the data-qubit state $\ket{s}$ ($s=0$ or $1$), the gate prepares a permutation-invariant distribution $p_n^{\ket{s}}$ over the number of excited ancillae, where $p_n^{\ket{s}}$ is the probability that exactly $n$ out of $N$ ancilla locations contain an excited atom.

Given a measurement record $\mathbf{m} = (m_1, \dots, m_N)$ and assuming that $n$ excited atoms are distributed uniformly among the $N$ locations, the likelihood is
\begin{equation}
\mathcal{L}(\mathbf{m}\mid n) = \frac{1}{\binom{N}{n}} \sum_{\substack{S \subseteq \{1,\dots,N\} \\ |S|=n}} \prod_{i\in S} P(m_i) \prod_{i\notin S} Q(m_i).
\end{equation}
This can be rewritten using the likelihood ratio $r_i = P(m_i)/Q(m_i)$ and the elementary symmetric polynomials $e_n(r_1,\dots,r_N)$:
\begin{equation}
\mathcal{L}(\mathbf{m}\mid n) = \frac{e_n(r_1,\dots,r_N)}{\binom{N}{n}} \prod_{i=1}^N Q(m_i).
\label{eq:likelihood_given_k}
\end{equation}
Marginalizing over the atom-number distribution yields the full likelihood of the measurement record under hypothesis $\ket{s}$:
\begin{equation}
\mathcal{L}(\mathbf{m}\mid \ket{s}) = \left(\prod_{i=1}^N Q(m_i)\right) \sum_{n=0}^{N} \frac{p_n^{\ket{s}}\; e_n(r_1,\dots,r_N)}{\binom{N}{n}}.
\label{eq:full_likelihood}
\end{equation}

\subsection{Results}
For each measurement record $\mathbf{m}$, we classify the data-qubit state by comparing log-likelihoods:
\begin{equation}
\hat{s} = \arg\max_{s\in\{0,1\}} \log \mathcal{L}(\mathbf{m}\mid \ket{s}).
\end{equation}
Note that this MLE classifier is \emph{optimal}, assuming perfect modeling of the noise.

\begin{figure*}
\centering
\includegraphics[width=\textwidth]{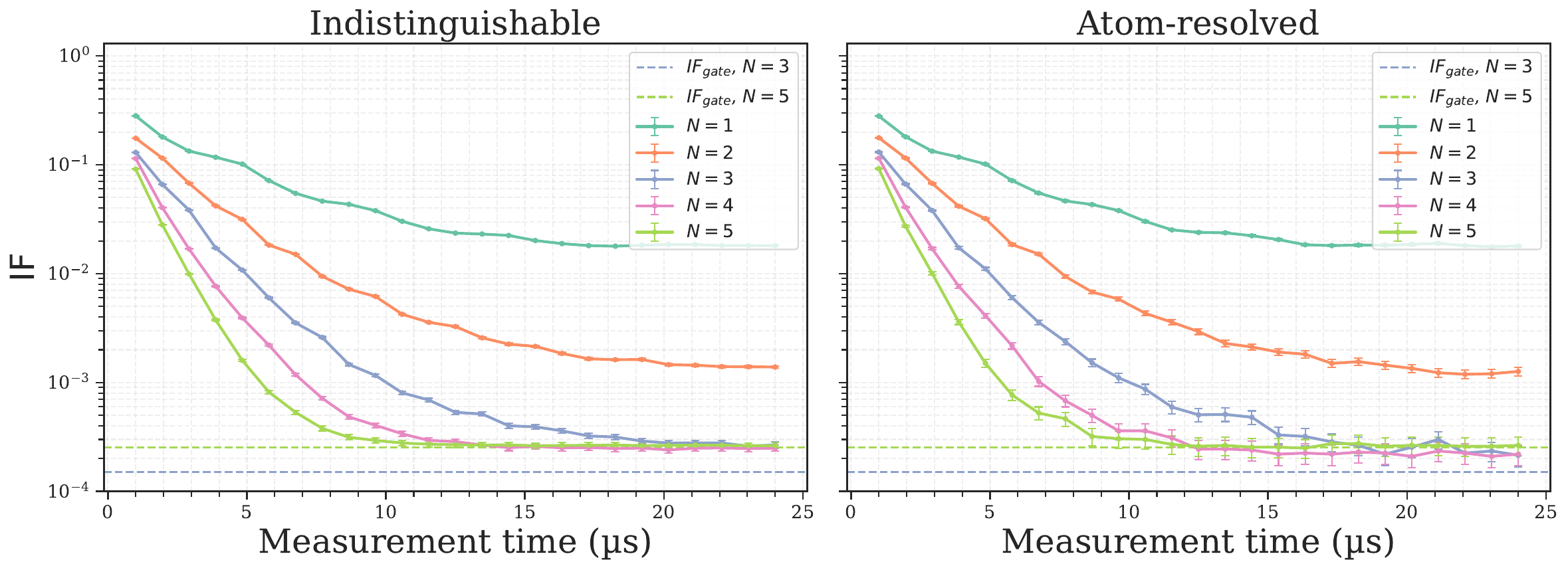}
\caption{Measurement infidelity as a function of measurement time for the indistinguishable (left) and atom-resolved (right) detection schemes. Dashed lines indicate the gate infidelity floor for $N=3$ and $N=5$. The two schemes yield nearly identical performance, indicating that atom-resolved detection offers no significant advantage for the parameters considered in this work.}
\label{fig:measurement_fidelity_atom_resolved}
\end{figure*}

The resulting measurement infidelity as a function of measurement time is shown in Fig.~\ref{fig:measurement_fidelity_atom_resolved} (right panel), alongside the indistinguishable (aggregated-count) model (left panel). We observe that the two schemes yield nearly identical performance across all ancilla numbers $N$ and measurement times. The small differences visible at longer measurement times are consistent with finite sampling noise, which is more pronounced in the atom-resolved case due to the higher computational cost of evaluating the  likelihood function. We conclude that, for the parameters considered in this work, atom-resolved detection offers no significant advantage over aggregated photon counting, and the simpler indistinguishable model is sufficient.

\section{Numerical Parameter}
\label{SI:numerical_table}
See table \ref{SI:numerical_table_params} for our numerical parameter.

\begin{table*}[t]
\centering
\caption{Numerical parameters used in the Cs--Rb simulation and in the photon-counting measurement model.}
\label{tab:sim_parameters}
\begin{tabular}{lll}
\toprule
\textbf{Quantity} & \textbf{Symbol} & \textbf{Value} \\
\multicolumn{3}{l}{\textit{Cs and Rb in $\ket{77S_{1/2}}$ and $\ket{78S_{1/2}}$ respectively}}\\
Cs--Cs vdW coefficient & $C_6^{\mathrm{Cs}}$ & $-2.9~\mathrm{GHz}\,\mu\mathrm{m}^6$ \\
Cs--Rb vdW coefficient & $C_6^{\mathrm{Cs\!-\!Rb}}$ & $-1.7~\mathrm{GHz}\,\mu\mathrm{m}^6$ \\
Cs radiative lifetime & $T_1^{\mathrm{Cs}}$ & $176~\mu\mathrm{s}$ \\
Rb radiative lifetime & $T_1^{\mathrm{Rb}}$ & $190~\mu\mathrm{s}$ \\
Minimum trap separation & $d$ & $2~\mu\mathrm{m}$ \\
\multicolumn{3}{l}{\textit{Photon-counting measurement model}}\\
Ancilla loss time & $T_{\mathrm{loss}}$ & $200~\mu\mathrm{s}$ \\
Fluorescence photon time & $T_{\mathrm{photon}}$ & $0.013~\mu\mathrm{s}$ \\
Background photon time & $T_{\mathrm{bg}}$ & $0.19~\mu\mathrm{s}$ \\
Camera solid angle & $\Omega_{\mathrm{cam}}$ & $0.1~\mathrm{sr}$ \\
Detection fraction & $\eta$ & $\Omega_{\mathrm{cam}}/(4\pi)=7.96\times 10^{-3}$ \\
\end{tabular}
\label{SI:numerical_table_params}
\end{table*}


\end{document}